# Design and Fabrication of a Novel Micro Electromagnetic Actuator

Chia-Yen Lee[1*], Zgen-Hui Chen[1], Hsien-Tseng Chang[2], Chiang-Ho Cheng[1], and Chih-Yung Wen[2]

[1]Department of Mechanical and Automation Engineering, Da-Yeh University, Changhua, Taiwan
[2]Department of Aeronautics and Astronautics, National Cheng Kung University, Tainan, Taiwan

*Abstract*- **A novel technique for the fabrication of micro electromagnetic actuators was proposed and a prototype was designed and fabricated in this study. The constituent parts of the designed actuator are comprised of a diaphragm, a micro coil, and a magnet. When an electrical current was applied to the micro coils, the magnetic force between the magnet and the coil was produced and caused the diaphragm to deflect and become the source of actuation. The structure of the actuating device used polyimide as the insulating barrier and electroplated copper as the coils. The diaphragm deflection can be regulated by varying the current passed through the micro coil and hence the actuating effects can be controlled. The results show that the maximum diaphragm deflection within elastic limits is 150 μm, obtained as applying a current of 0.6 A through the micro coil with 100 μm line width. The micro electromagnetic actuator proposed in this study is easily fabricated and is readily integrated with existing bio-medical chips due to its plane structure.**

## I. Introduction

In the twenty-first century, both environmental protection and economic development have become two major considerations regarding consumer products. Forms of the products tend to be light, thin, short and compact, and their functions consider space, energy, and materials reduction. In other words, there should be additional cost in the fabrication of the products, which is the major purpose for the process technology development [1]. Micro-Electro-Mechanical Systems are regarded as a major technology and may become a proactive research field in the future. Micro-actuators are machines which need input controlled signals, such as voltage and electric current. These signals must be converted into kinetic energies for movement such as horizontal movement, vertical vibration, and swinging. Moreover, their component size should be in the micron level [2]. Static, electromagnetic, and piezoelectric [3-10] actuators are the principal devices in the Micro-Electro-Mechanical System. Liu *et al.* [3] fabricated magnetic flaps supposed to be as an integrated part of an active micro electro-mechanical fluid control system. Though a large deflection ( ~ 100 μm ) was achieved, the integration into a micro-fluidic system was not demonstrated in their study. Judy and Muller [4] proposed a magnetically activated device for switching light paths in microphotonic systems, which was still hard to be integrated into planar biochips as an actuator with a pumping function. Lagorce *et al.* [5] presented magnetic microactuators based on polymer magnets with good simulated and experimental results. Their cantilever design with small defection (~ 20 μm) was not demonstrated as a good pumping element in microfluidic systems. A valveless micropump using electromagnetic actuation was successfully presented [6]. A PDMS (polydimethylsiloxane) membrane with a magnet was actuated by an externally electromagnetic actuation. A successful pumping phenomenon was observed with the developed micropump in spite a bulky and external actuation device was required.

The present study presents a new micro electromagnetic actuator utilizing a PDMS membrane with a magnet. The actuator is integrated with micro coils to electromagnetically actuate the membrane and results in a large deflection with 150 μm . The micro electromagnetic actuator proposed in the study is easily fabricated and is readily integrated with existing bio-medical chips due to its plane structure.

## II. DESIGN

As shown in Fig.1, the actuator was constructed with the micro coil electroplated onto the glass substrate and the PDMS diaphragm [11] with permanent magnets bonded to the PMMA layer. The copper coil was electroplated onto the glass substrate using a standard surface micro-machining process. The PDMS diaphragm was fabricated by spin-coating the PDMS materials onto the glass wafer. The magnet used for the actuator is a permanent magnet. Three types of micro coils were designed: type I – 100 μm line width with 80 μm spacing (see Fig 2), type II – 200 μm line width with 60 μm spacing (see Fig 3), and type III – 400 μm line width with 320 μm spacing (see Fig 4). The size of the selected magnet was 1.5*1.5 mm with a magnetization retentive capacity of 2,300 Gauss.

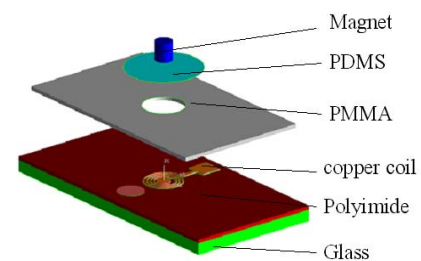

Figure 1 Sketch for the construction of the fabricated electromagnetic actuator.

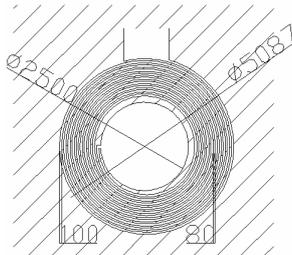

Figure 2 Type I micro coil design. The width, spacing and thickness of the coil are 100 μm, 80 μm and 20 μm, respectively.

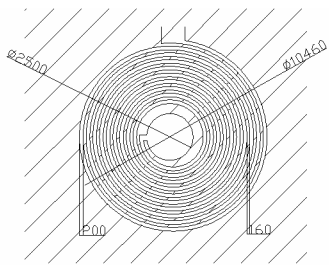

Figure 3 Type II micro coil design. The width, spacing and thickness of the coil are 200 μm, 160 μm and 20 μm, respectively.

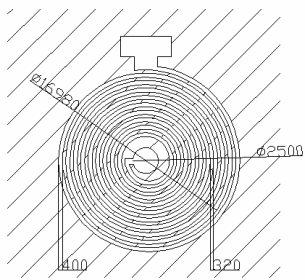

Figure 4 Type III micro coil design. The width, spacing and thickness of the coil are 400 μm, 320 μm and 20 μm, respectively.

### III. Fabrication

The prototype of the novel electromagnetic actuator was fabricated by surface micro-machining technology which includes the standard photo lithography, vacuum evaporation and the electroplating processes. Micro spiral coils are first fabricated using standard photo lithography and electroplating processes. A simplified fabrication process of the micro coils is shown in Fig 5. The fabricated spiral micro coils were designed to be 10 turns and the process started with a glass substrate. The fabrication process of the micro coils is described as follows:

(a) Onto the glass substrate a seed layer consisting of 30 nm chromium and 150 nm copper was deposited using electron-beam evaporation.
(b) A 15 μm thick of polyimide layer (PW-1000, Toray Industries Inc., Japan) was spinned on the seed layer with a via hole to connect the seed layer and the micro coil. Then the substrate was cured in a 350°C of oven to solidify the polyimide layer.
(c) Electroplating the via hole with 15 μm thick of copper.
(d) A 25 μm thick of photo resist layer (AZ 4620, Clariant Corp., Switzerland) was then deposited on the top of the polyimide layer and patterned to form molds of the micro coils. A seed layer of 100 nm copper was deposited using electron-beam evaporation.
(e) The copper coils were electroplated for 20 μm of thickness. Upon completion of the electroplating, the photo resistant mold was removed with acetone.
(f) A 15 μm thick of polyimide layer served as an insulation layer was spinned on the top of the copper coils and cured at 350°C for 1 hour after removing the photoresist. Finally, drilled holes in the polyimide mold allowed the connection of the coil.

The photographs of the prototypes of the three designed micro coils were shown in Figs. 6, respectively. Fig. 7 represents photographs of the fabricated PDMS diaphragm with a thickness of 100 μm and the bonded part of the PDMS diaphragm, the PMMA layer, and the magnet. A hole was drilled through the PMMA layer with a diameter of 4mm in its center. A photograph of the fabricated magnetic actuator is represented in Fig. 8. The parameters of the components of the fabricated actuator are summarized in tables 1-3.

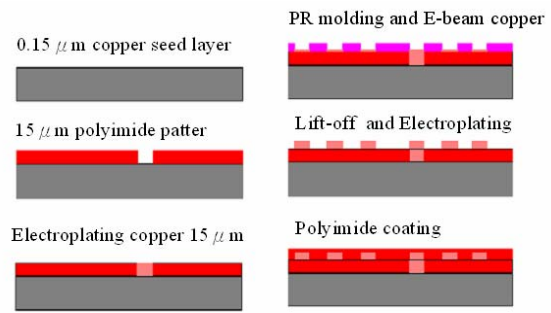

Figure 5 Simplified fabrication process for the micro coils.

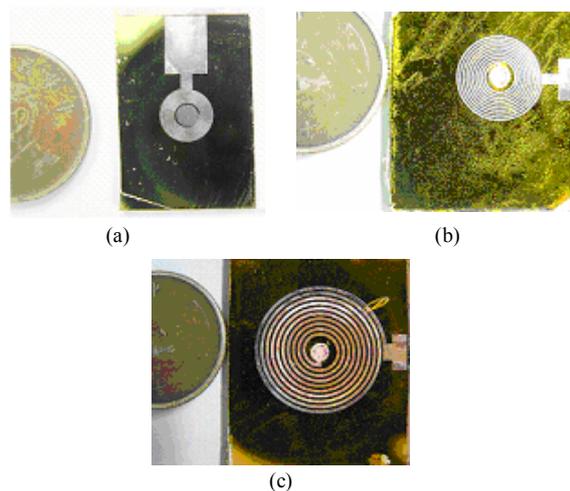

Figure 6 Micro coil prototype design of (a) type I, (b) type II and (c) type III.

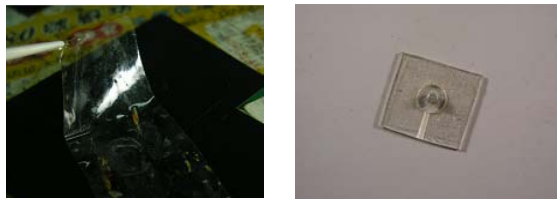

(a)           (b)

Figure 7 Photographs of (a) the fabricated PDMS diaphragm with the thickness of 100 μm and (b) the bonded part of the PDMS diaphragm, the PMMA layer and the magnet.

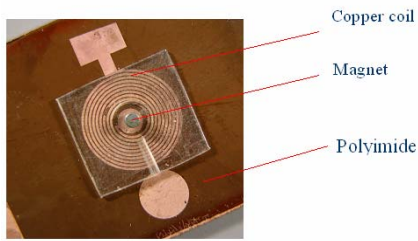

Figure 8 Fabricated electromagnetic actuator in the research.

Table 1 Parameters of the fabricated diaphragm.

| Parameter | Material | Radius | Thickness |
|---|---|---|---|
| Data | PDMS | 2000 μm | 100 μm |

Table 2 Parameters of the magnet.

| Parameter | Material | Radius | Thickness | Remanence(Br) |
|---|---|---|---|---|
| Data | NdFeB | 750 μm | 1500 μm | 2.3 kG |

Table 3 Parameters of the fabricated micro coil.

| Parameters | Material | Width | Pitch | Spacing | Turns | Inner radius | Outer radius | Resistance (at 20 ℃) |
|---|---|---|---|---|---|---|---|---|
| Data | Cu | 100 μm | 20 μm | 80 μm | 10 | 2500 μm | 5087 μm | 2 Ω |

## IV. EXPERIMENTAL RESULTS

In figure 9, the magnetic field produced by the micro coils was measured using a Tesla meter (model 4048, F. W. Bell, U.S.A.). The variation of the flux density with the vertical distance was measured along the central axis of the coil from the coil center for an input current of 0.5 A. Fig. 10 compares the experimental results of the three micro coils. Furthermore, the corresponding derivatives of the flux density, as shown in Fig. 11, which shows the gradient of the magnetic field produced by these three coils. It can be seen that the maximum gradient of the magnetic field occurs at a point located at 500 μm above the planar coil. Therefore, the magnet should be located at this position to optimize the electromagnetic actuation effect and the greatest gradient of the magnetic field was generated by the coil with type I micro coil.

Fig. 12 shows the experimental setup for the diaphragm displacement measurement (power supply: PR8323, ABM, Taiwan / laser displacementmeter: LC-2400A + 2430, Keyence, Japan). The displacement of the diaphragm was measured at its central area, which is occupied by the magnet, where the maximum deflection occurs in the diaphragm. The diaphragm displacement generated by these three coils is shown in Fig. 13. It can be found that the greater deflection was produced with type I micro coil and the maximum deflection within elastic limit was found to be 150 μm as applying a current of 0.6 A.

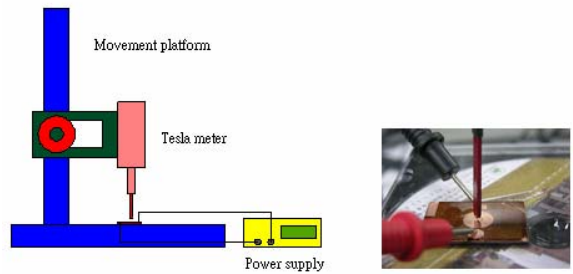

Figure 9 Experimental setup for magnetic flux density measurement with different vertical distance.

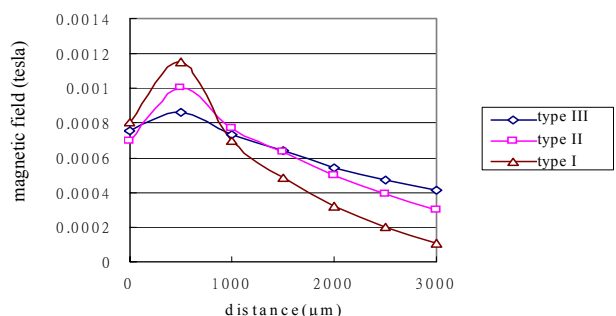

Figure 10 Magnetic field produced by the micro coils.

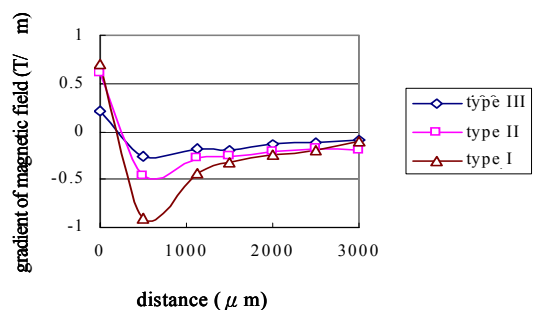

Figure 11 Gradient of the magnetic field.

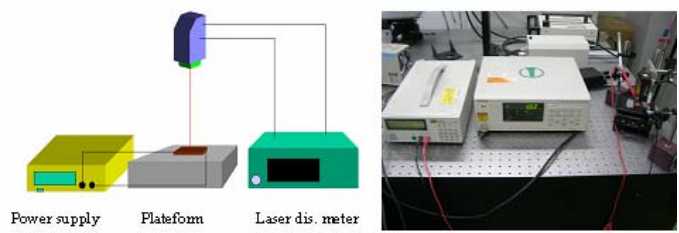

Figure 12 Experimental setup for diaphragm displacement measurement.

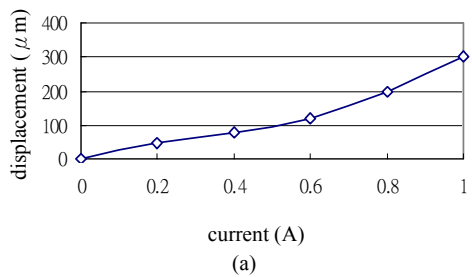

(a)

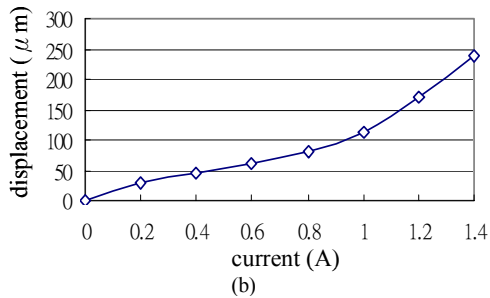

(b)

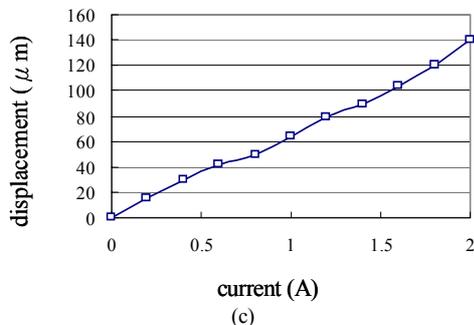

(c)

Figure 13 Membrane displacement of micro coils (a) type I, (b) type II and (c) type III at different currents.

## V. CONCLUSIONS

The magnetic actuator utilizing PDMS diaphragm was designed and fabricated in the current study. In the experiments, the fabricated actuator exhibited excellent actuation response with large diaphragm deflections. The characteristics of the fabricated actuator make the device useful for micro pumps and other actuator applications. The results of this study provide a valuable contribution to the ongoing development of Lab-on-a Chip systems due to its planar structure.

## ACKNOWLEDGMENT

The authors would like to thank the financial supports provided by the National Science Council in Taiwan (NSC 96-2221-E-212-037 and NSC 95-2218-E-006-022).

## REFERENCES

[1] U. Neda, K. Nakmura and T. Takumi: Sensors and Actuators A 54 (1996) 626.
[2] E. Kalvesten: IEEE MEMS 98 (1998) 574.
[3] Liu, T. Tsao, Y. C. Tai and C. M. Ho: IEEE MEMS 94 (1994) 57.
[4] J. W. Judy and R. S. Muller: IEEE J. of MEMS 6(3) (1997) 249.
[5] L. K. Lagorce, O. Brand and M. G. Allen: IEEE J. of MEMS 8(1) (1999) 2.
[6] C. Yamahata, C. Lotto, E. Al-Assaf and M. A. M. Gijs: Microfluid Nanofluid 1 (2005) 197.
[7] K. Kurihara, M. Hida, S. Umemiya and S. Koganezawa: Jpn. J. Appl. Phys. 43 (2004) 6725.
[8] K. Kurihara, M. Hida, S. Umemiya, M. Kondo and S. Koganezawa: Jpn. J. Appl. Phys. 45 (2006) 7471.
[9] Y. Sakai, T. Futakuchi, T. Iijima and M. Adachi: Jpn. J. Appl. Phys. 44 (2005) 3099.
[10] K. Tanaka, T. Konishi, M. Ide, Z. Meng and S. Sugiyama: Jpn. J. Appl. Phys. 44 (2005) 7068.
[11] J. C. Yoo, M. C. Moon, C. J. Kang, D. Jeon and Y. S. Kim: Jpn. J. Appl. Phys. 45 (2006) 519.